# Single-layer graphene on epitaxial FeRh thin films

Vojtěch Uhlíř[a,b,*], Federico Pressacco[c], Jon Ander Arregi[a], Pavel Procházka[a,b], Stanislav Průša[a,b], Michal Potoček[a,b], Tomáš Šikola[a,b], Jan Čechal[a,b], Azzedine Bendounan[d] and Fausto Sirotti[d,e]

[a]CEITEC BUT, Brno University of Technology, Purkyňova 123, 612 00 Brno, Czech Republic

[b]Institute of Physical Engineering, Brno University of Technology, Technická 2, 61669 Brno, Czech Republic

[c]Center for Free Electron Laser Science, University of Hamburg, Luruper Chaussee 149, 22761 Hamburg, Germany

[d]Synchrotron SOLEIL, Saint-Aubin, BP 48, F-91192 Gif-sur-Yvette Cedex, France

[e]Physique de la Matière Condensée, CNRS and École Polytechnique, Université Paris Saclay, F-91128 Palaiseau, France

*Email: vojtech.uhlir@ceitec.vutbr.cz

Graphene is a 2D material that displays excellent electronic transport properties with prospective applications in many fields. Inducing and controlling magnetism in the graphene layer, for instance by proximity of magnetic materials, may enable its utilization in spintronic devices. This paper presents fabrication and detailed characterization of single-layer graphene formed on the surface of epitaxial FeRh thin films. The magnetic state of the FeRh surface can be controlled by temperature, magnetic field or strain due to interconnected order parameters. Characterization of graphene layers by X-ray Photoemission and X-ray Absorption Spectroscopy, Low-Energy Ion Scattering, Scanning Tunneling Microscopy, and Low-Energy Electron Microscopy shows that graphene is single-layer, polycrystalline and covers more than 97% of the substrate. Graphene displays several preferential orientations on the FeRh(001) surface with unit vectors of graphene rotated by 30°, 15°, 11°, and 19° with respect to FeRh substrate unit vectors. In addition, the graphene layer is capable to protect the films from oxidation when exposed to air for several months. Therefore, it can be also used as a protective layer during fabrication of magnetic elements or as an atomically thin spacer, which enables incorporation of switchable magnetic layers within stacks of 2D materials in advanced devices.

## Introduction

Interfaces of magnetic thin films with graphene offer a rich realm of functionalities that play a key role in a number of magnetic and spintronic phenomena [1-8]. Graphene plays two major roles in prospective spintronic devices: it represents an active material, exhibiting or modifying magnetic properties, and acts as a stabilizing and protective layer [9]. In this respect, the use of 2D materials to separate reactive films from air contaminants is increasingly realized [9-13], with a possibility to achieve efficient prevention from oxidation even in corrosive environments [14]. Regarding magnetism, modification of magnetic properties by graphene was demonstrated in enhancement of the perpendicular magnetic anisotropy [2,15] and spin reorientation transition in Co thin films [16]. Recently, chiral magnetic structures were found in ultrathin Co and Ni films due to the Dzyaloshinskii-Moriya interaction at the graphene/ferromagnet interface [3]. Graphene has also been shown to mediate interlayer exchange coupling between magnetic layers [17]. Graphene electronic and magnetic properties are affected by the presence of a ferromagnetic (FM) film as well: besides very long spin diffusion length [18], induction of magnetism in graphene was observed for instance in graphene/Ni [19] and graphene/Co [20] systems. Recently, it has been predicted that magnetic properties of graphene can be tuned by electric polarization in multiferroics due to the proximity effect [21]. The realization of graphene layers on magnetically transformable materials would bring another level of external control to spintronic devices.

Here we focus on graphene formation on epitaxial FeRh thin films which feature a first-order antiferromagnetic (AF) to ferromagnetic phase transition present around 360 K [22]. Due to the interconnected structural, electronic and magnetic order parameters, FeRh is a highly tunable material [23,24] where the phase transition can be controlled by different stimuli: temperature, magnetic field, strain, electrical current and optical pulses [23,25-29]. The phase transition is accompanied by a modification of the FeRh lattice parameter of about 1% [30]. Accordingly, small variations of the lattice parameter can induce the phase transition: strain relaxation at the Rh-terminated surface of a film in the AF phase is sufficient for formation of a double FM layer [31]. Sensitivity to the modification of lattice parameter or structure size [32,33] opens the way to a large number of applications. Electric field control of magnetic order above room temperature has been obtained by deposition of FeRh on ferroelectric $BaTiO_3$ [25]. Controllable giant magnetization changes are induced by a structural phase transition of FeRh in a metamagnetic artificial multiferroic [34]. The FeRh system has also been the object of several studies of ultrafast phase transition dynamics driven by femtosecond laser pulses. Ultrafast modification of physical properties can be observed in all the order parameters: magnetic [35], structural [28], and electronic [29]. Moreover, the temperature dependent phase transition has been used to produce a room temperature bistable AF memory insensitive to strong magnetic fields and producing negligible stray fields [36], thus presenting a platform that may be exploited in AF spintronics [37].

In the following we present a detailed characterization of the graphene/FeRh system, which features good epitaxy and strong attachment of graphene to a highly tunable magnetic substrate promising new ways of controlling graphene properties.

**Experimental**

Graphene was prepared following two alternative approaches: utilizing carbon segregation from a FeRh film at elevated temperatures and by decomposition of ethylene on a clean FeRh surface as discussed below. Epitaxial FeRh thin films used as substrates for graphene growth were prepared on MgO(001) substrates by dc magnetron sputtering using an equiatomic target. The films were grown at 725 K and post-annealed in situ at 1070 K for 45 minutes at a residual pressure of $10^{-7}$ mbar in the magnetron vacuum chamber. For comparison, we prepared 55-nm-thick and 150-nm-thick FeRh films. All samples show a hysteretic behavior of magnetization as a function of temperature, which is a signature of the AF to FM transition (Fig. S1 in the Supplementary Material).

In case of graphene growth using carbon segregation from the FeRh film, graphene forms on the surface as long as the carbon content in the FeRh substrate is sufficient. We have found that the carbon reservoir in 55-nm-thick films was not enough to form a graphene layer, whereas in 150-nm-thick films it was. The general process of carbon segregation towards the FeRh surface upon annealing is demonstrated by elemental depth profiling using Secondary Ion Mass Spectroscopy (SIMS, see Fig. S2).

Removing the graphene layer by mild Ar sputtering with subsequent annealing of the sample leads to a gradually smaller graphene coverage of the substrate and further sputtering/annealing cycles lead to a clean FeRh surface due to depletion of the carbon supplies. This approach was taken for the 150-nm-thick FeRh films prior to the X-ray Photoemission Spectroscopy (XPS) experiments by low-energy (500 eV) Ar sputtering in the TEMPO beam line preparation chamber at Synchrotron Soleil. The sample was then annealed at 950 K for 20 minutes in a pressure lower than $2\times10^{-7}$ mbar, which led to formation of single-layer graphene. Further, total three sputtering and annealing cycles were performed to obtain a clean Rh-terminated surface [31].

The second approach for graphene growth is based on decomposition of ethylene on a heated FeRh surface. For this purpose, we have used the 55-nm-thick FeRh film that was exposed to ethylene for 120 s at 910 K and $2\times10^{-7}$ mbar in a Low-Energy Electron Microscopy (LEEM) chamber, which enabled in-situ observation of the graphene formation (see Fig. 1). This sample was prepared 8 months before the synchrotron x-ray spectroscopy experiments. Prior to the x-ray analysis, the sample was heated to 770 K for 30 min to clean the surface from airborne contamination. Both ways of graphene production led to graphene with similar properties: most measured characteristics did not show significant differences.

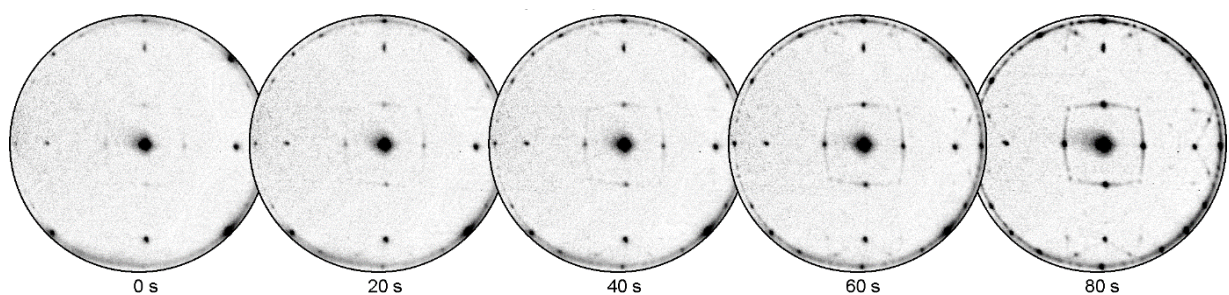

**Fig. 1.** LEED diffraction pattern of a FeRh surface exposed to ethylene for a total time of 120 s at a pressure of $2\times10^{-7}$ mbar and temperature 910 K. Graphene diffraction peaks are visible almost immediately upon ethylene exposure and become dominant in the later stages of the growth. Complete identification of the diffraction spots is detailed in Fig. S3.

After the x-ray spectroscopy analysis at Synchrotron Soleil, the samples were extracted to atmosphere for 1 month and transferred back to ultra-high vacuum (UHV) experimental stations at CEITEC BUT. The UHV complex system was used to perform in-situ analysis by Scanning Tunneling Microscopy (STM), Low Energy Ion Scattering (LEIS), and LEEM techniques without breaking the UHV. Prior to the STM, LEEM and LEIS experiments, the samples were heated to 500 K for 30 min to clean the surface from airborne contamination.

**Results and Discussion**

The surface composition and electronic properties of the graphene layers were analyzed using XPS and Near-Edge X-ray Absorption Fine Structure (NEXAFS) at the TEMPO beamline at the SOLEIL synchrotron radiation facility. The XPS overview spectra measured on a clean FeRh sample and sample with a graphene layer are compared in Fig. 2a. The photon energy is chosen such that the kinetic energy range of photoelectrons covers energy levels of all the elemental components to be unambiguously identified. After the repeated cleaning procedure, only photoemission and Auger peaks associated with Rh and Fe atoms are identified and no contamination from carbon and oxygen atoms is observed.

A detail of the Rh 3d and C 1s peaks (Fig. 2b) indicates that carbon is only present after the first sputtering/annealing cycle. No oxygen was detected after any of the sputtering/annealing cycles. Furthermore, the XPS analysis showed no oxygen contamination in any of the graphene-covered samples. Fig. 2c shows that the surface stoichiometry is not modified by the sputtering/annealing processes, or by the presence of graphene: quantitative analysis of the Fe 3p and Rh 4p peak areas confirms that the FeRh thin film is terminated by a Rh atomic plane [31].

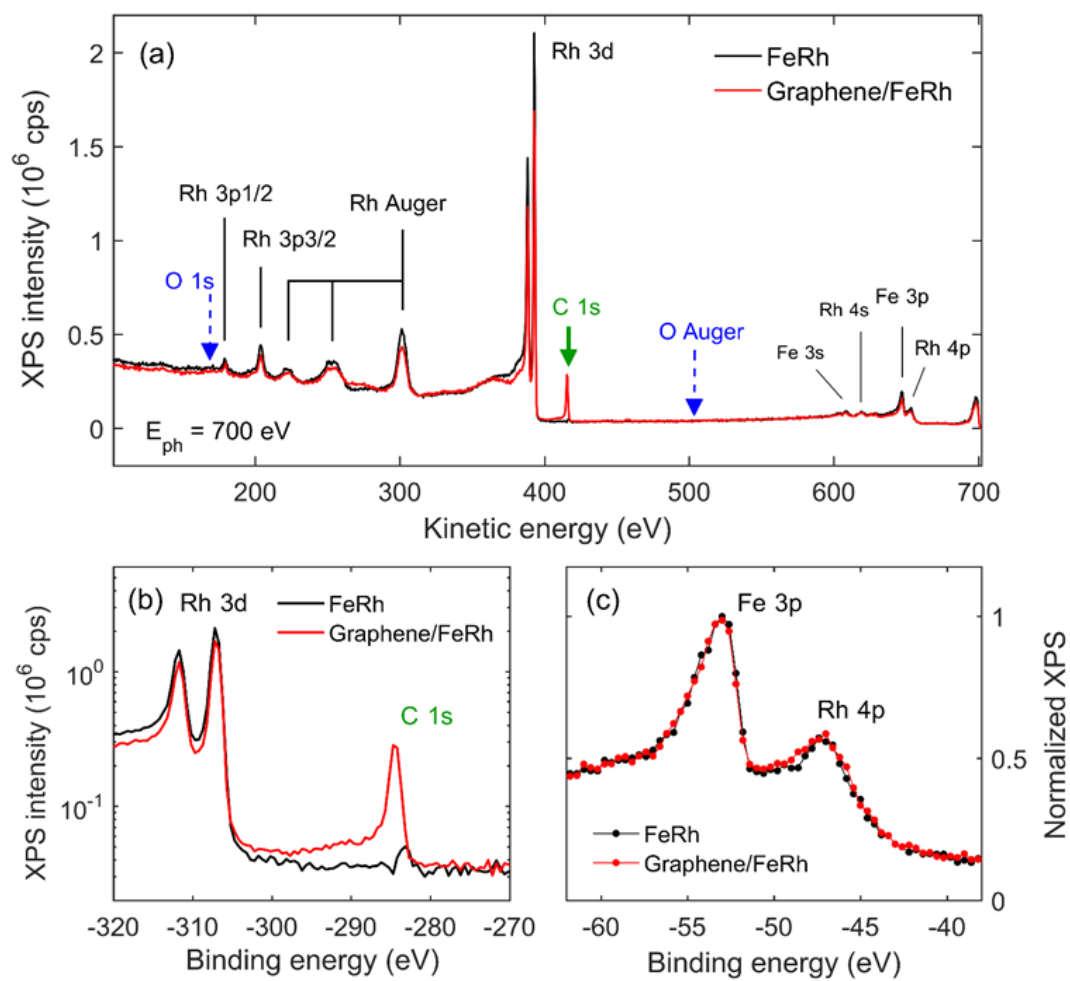

**Fig. 2.** (a) Wide-energy-range XPS spectra measured using 700-eV photons for the graphene-covered surface (after the first sputtering/annealing cycle, red line) and clean surface without graphene (after two additional sputtering/annealing cycles, black line). The peaks are labeled either Auger or via the corresponding core level. The Fe and Rh photoemission and Auger peaks along with the C 1s and O 1s photoemission peaks are indicated by black lines, and green and blue arrows, respectively. (b) A detail of the photoemission spectra covering the Rh 3d and C 1s peaks in the logarithmic scale. (c) Comparison of the Fe 3p and Rh 4p photoemission peaks obtained for the clean surface (black line) and graphene-covered surface (red line), presenting identical Fe-to-Rh stoichiometry.

Quantification of the carbon contribution to the surface composition is done by considering the electron inelastic mean free path of 0.67 nm [38], the C 1s and Rh 3d photoionization cross-sections ratio of 0.09 [39], and using the same modeling approach as we applied in ref. 31. For this purpose, a detail of the energy range comparing the C 1s photoemission signal with that of Rh 3d core levels is shown in Fig. 2b. Assuming the graphene layer present on top of the FeRh surface, the relative C 1s peak intensity corresponds to 0.8-1.2 atomic layers. Repeating the sputtering/annealing process on different samples resulted in the same relative intensity of the C 1s core level.

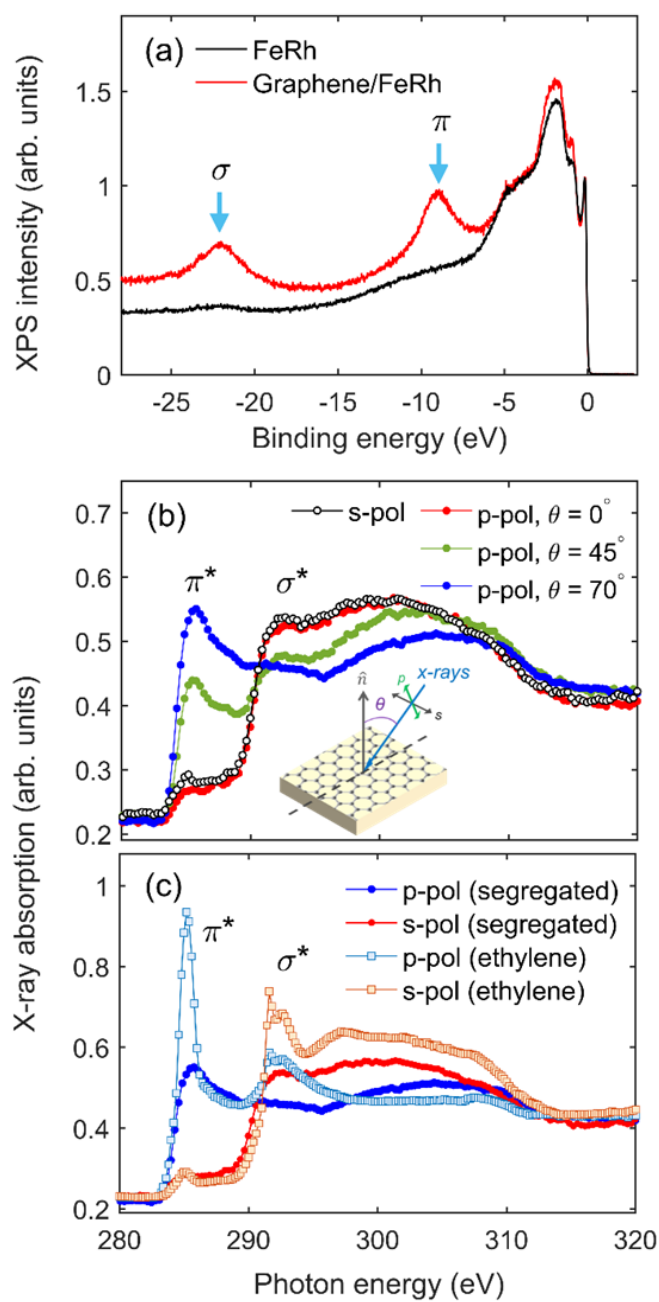

**Fig. 3.** (a) Valence band photoemission spectra measured along the $\Gamma$ direction (normal emission) of the FeRh epitaxial layer with a clean surface (red line) and graphene-protected surface (black line). The $\sigma$ (22 eV) and $\pi$ (10 eV) bands of graphene are indicated. (b) Linear dichroism at the C 1s absorption edge confirming $sp^2$ hybridized graphene orbitals on the sample surface. Absorption of *p*-polarized x-rays exciting electrons to the $\pi^*$ state increases with the angle of incidence $\theta$ (see the inset). At 0° (normal incidence) the x-ray polarization is orthogonal to graphene $\pi$ orbitals and the electrons are preferentially excited to the $\sigma^*$ state, which is equivalent to the case with incident *s*-polarized x-rays. (c) Comparison of the x-ray linear dichroism at the C 1s edge of graphene segregated from the FeRh layer (filled circles) and grown under ethylene exposure (open squares).

Electronic properties of the graphene layer were analyzed by valence band spectroscopy. The photoemission spectra measured at the Γ point for the samples with and without the graphene layer are compared in Fig. 3a. The two broad structures at 10-eV and 22-eV binding energies can be identified as graphene π and σ bands, respectively [43]. The graphene structure is further confirmed by measuring linear dichroism at the carbon K edge (Fig. 3b). The arrangement of carbon atoms in graphene and the resulting sp$^2$ hybridization makes the x-ray absorption process strongly dependent on the orientation of the electric vector of the x-rays with respect to the surface. Therefore, the probability of electron excitation from the K-shell to π* or σ* states depends on the orientation of the incident radiation polarization vector with respect to the graphene basal plane. The results obtained on the FeRh layer after the first annealing process are presented in Fig. 3b. The high contrast in the linear dichroism observed at the π* state as a function of the angle of incidence $\theta$ is a well-established spectral fingerprint of the sp$^2$ hybridized carbon orbitals, which indicates presence of a graphene layer on the surface [1,44]. Although both graphene prepared by carbon segregation and graphene from ethylene generally displayed very similar characteristics, in case of x-ray linear dichroism we observe more prominent peaks at the C 1s edge for the ethylene graphene (Fig. 3c).

In order to get a more precise determination of graphene coverage and to characterize homogeneity of the graphene layer, we have used LEIS. LEIS is known for its extreme surface sensitivity and possibility to provide quantitative composition of the outermost surface layer [40]. It has also proved its ability in detailed characterization of graphene on silica substrate [41]. We have used a 3-keV He ion beam at a scattering angle of 145° to obtain a spectrum of the FeRh sample covered by graphene (Fig. 4). The kinetic energy of the He projectile after the collision identifies the atomic mass of the target atoms following rules of elastic binary collisions. The carbon peak is located at an energy of 830 eV. The surface peaks of Fe and Rh are low in intensity even though the differential scattering cross-sections [42] of Fe and Rh are larger than the carbon one by a factor of 10 and 19, respectively. It is evident that the graphene layer covers a significant part of the analyzed surface.

Regarding quantification of the graphene coverage, it is more precise to determine the complementary uncovered surface area, as the carbon peak intensity is influenced by the character of the local carbon bonds (sp$^2$ vs. sp$^3$ hybridization [41]). For this purpose, calibration measurements for He-ion scattering on pure Fe (red spectrum) and Rh (blue spectrum) reference samples were performed. The signal intensities (peak areas) of Fe and Rh measured at the analyzed surface are compared to the signal intensities of Fe and Rh in elemental reference samples (see Fig. 4). The ratio of the Rh peak intensities is $88/14780=6.0\times10^{-3}$, i.e. about 0.6% of the analyzed surface corresponds to Rh. Similar evaluation of Fe data gives the ratio of $50/9308=5.4\times10^{-3}$, hence 0.5% of the surface corresponds to Fe.

The ratio of atomic surface densities of elements in the areas not covered by graphene can be quantified by considering the atomic surface densities of reference polycrystalline Fe ($1.93\times10^{15}$ atoms/cm$^2$) and Rh ($1.74\times10^{15}$ atoms/cm$^2$) calculated from their mass densities. The corresponding ratio of atomic surface densities of Fe and Rh is then 1.07 (experimental error 5%). Interestingly, we obtain this Fe to Rh ratio close to 1 for both the graphene-free sputtered (disordered) and annealed (reconstructed) FeRh surface. In the latter case, even a clean Rh-terminated surface exposes Fe atoms which are located in the second topmost monolayer below the centers of the square Rh cells [31] and thus are accessible to the impinging He ions. Although this fact would lead to overestimation of the FeRh surface area when simply summing the Fe and Rh surface coverages determined above, we can still use the sum to obtain a lower bound of the complementary graphene coverage. Hence, it amounts to about 98% of the sample surface area, considering that there is no other element detected except of carbon. It is evident that the relative error of the Fe and Rh signal evaluation from the tiny peaks in the spectrum (black trace) is significant and amounts to about one third of the values presented in the plot. This leads to correction of the lower bound of the graphene coverage down to 97%.

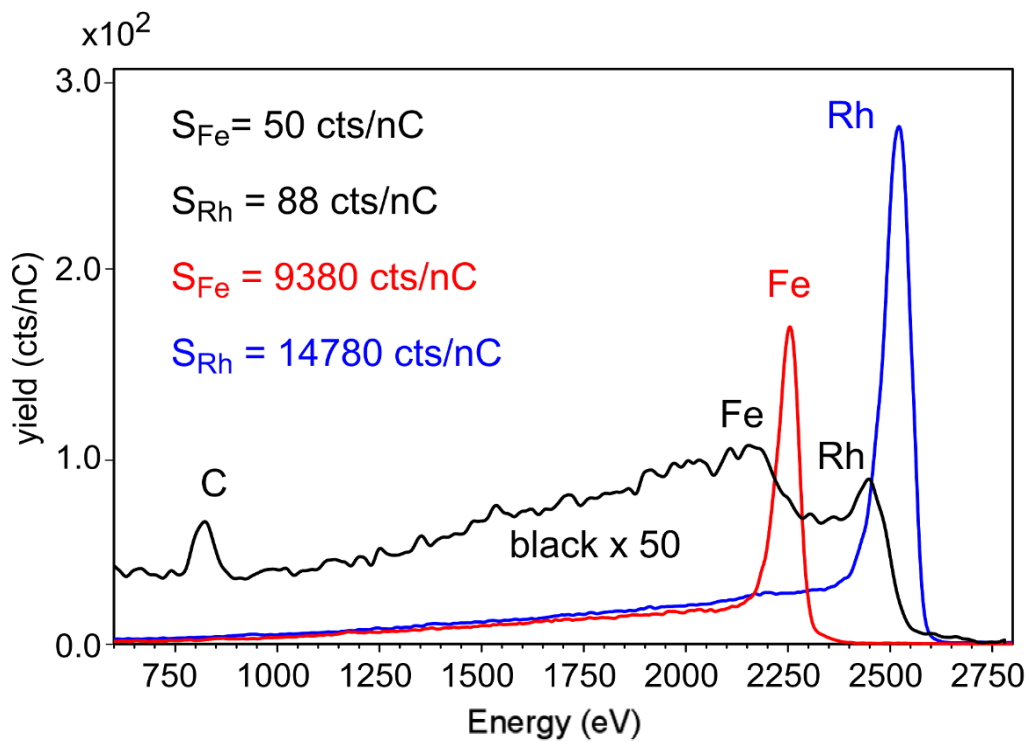

**Fig. 4.** LEIS spectra measured at the graphene-covered sample surface (black) and at pure Fe (red) and Rh (blue) reference samples using identical experimental conditions (He 3 keV, scattering angle 145°). The black spectrum was multiplied by a factor of 50. The numeric values $S$ represent the intensity of the particular peaks in the given spectrum after background subtraction. The ratio of $S$ for the analyzed sample and the corresponding reference gives surface coverage of the particular element.

The in-plane orientation of the graphene layer was determined using STM. STM images of the graphene layer on FeRh(001) with atomic resolution are presented in Fig. 5a and display several domains with distinct atomic-like contrast. Except for the localized defects discussed below, the substrate is atomically flat, which together with the observations from LEIS and XPS suggests that monolayer graphene covers most of the surface. The graphene layer is broken up into domains that are contiguous and no graphene flake edges are apparent. As the substrate is an epitaxial FeRh film on a single-crystalline MgO (see x-ray diffraction characterization in ref. 32) the domains can be associated with different orientations of graphene and display a moiré-like structure of the graphene hexagonal lattice superimposed on the underlying square lattice of the FeRh substrate.

Due to the relative crystallographic orientations of graphene domains and the substrate, the moiré-like contrast reminds the quasi-hexagonal reconstruction on Pt(001) [45] rather than the contrast typically observed for graphene on Rh(111) [46,47]. However, using a simple model of overlapping the graphene lattice over the square FeRh lattice it is not straightforward to unambiguously identify the moiré patterns and DFT calculation would be needed to reproduce the contrast.

Absence of apparent wrinkles indicates strong attachment of graphene to the surface. At some sites the graphene is delaminated and appears like quasi-free-standing as shown in Fig. 5b-d. FFT in Fig. 5e-g shows that these graphene nanobubbles possess different orientations, which is also expressed by a distinct appearance and rotation of the line moiré pattern in the nanobubble surroundings. Such nanobubbles are typically formed to relax the strain caused by strong attachment to the surface and can be a source of unique magnetic behavior [48].

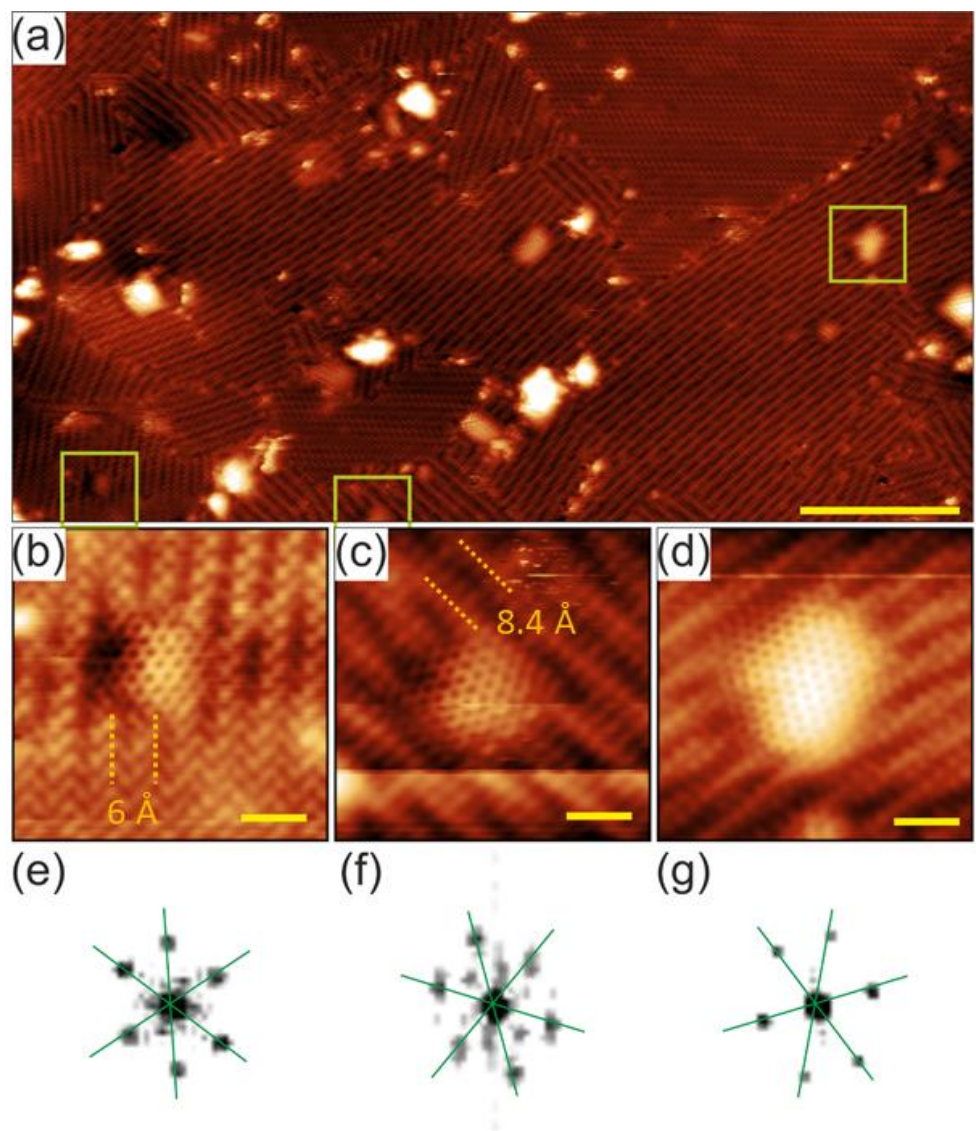

**Fig. 5.** (a) Wide field image of the graphene/FeRh surface taken by STM. Distinct atomic-like patterns in several domains are clearly visible. (b)-(d) Magnified view of areas marked in (a) by squares show parts of graphene delaminated from the substrate displaying the characteristic hexagonal structure. The periodicity of the moiré pattern is indicated by dotted lines. (e)-(g) FFT taken over the delaminated graphene nanobubbles corresponding to (b)-(d) shows distinct orientations of graphene in individual domains. The rotation of the diffraction pattern with respect to (e) is +15° in (f) and −11° in (g). Scale bars: 10 nm in (a) and 1 nm in (b)-(d).

An overall description of the orientations of graphene domains with respect to the FeRh lattice was obtained by Low-Energy Electron Diffraction (LEED) from an area of 10×15 µm$^2$ using an electron energy of 50 eV (Fig. 6a). The diffraction pattern looks very complex featuring lines in addition to sharp diffraction spots. As discussed in more detail in the Supplementary Material, the ring close to the edge of the diffraction pattern (Ewald's sphere projection) in Fig. 6a is a clear sign of a polycrystalline graphene layer with many different domain orientations. In addition, we recognize several preferential graphene orientations giving sharp spots. The most prominent one is marked by red and blue dots in the diffraction model shown in Fig. 6b. Here, the rotation of the graphene reciprocal unit vector $\vec{a}_G^*$ with respect to the FeRh reciprocal unit vector $\vec{a}_{FeRh}^*$ is 0°. In real space the corresponding rotation of the graphene unit vector $\vec{a}_G$ with respect to the FeRh unit vector $\vec{a}_{FeRh}$ is ±30°. The spots associated with the second most prominent orientation are rotated by ±45° corresponding to real-space graphene rotation by ±15°. The remaining distinguishable diffraction spots are associated with graphene real-space rotations by ±11° and ±19°. The line features are associated with the moiré effect originating from all possible rotations of graphene with respect to the FeRh substrate, whereas the sharp spots on these lines are moiré spots of the preferential orientations.

The bright field LEEM image showing an incomplete coverage of the surface by graphene (light regions in Fig. 6c) was captured after 80 s during graphene layer growth by ethylene exposure (120 s in total, see Section 2). The dark field LEEM images (Fig. 6d) show the corresponding spatial distribution of domains with preferential graphene orientations (±30°), obtained by selecting the related moiré diffraction spots (blue and red circles in Fig. 6a).

Diffraction patterns (µ-diffractions) in Fig. 6e-f (left) are collected from a circular area with a diameter of 185 nm on the FeRh surface using a µ-diffraction aperture. The patterns were measured at two different positions on the sample and represent graphene domains larger than 185 nm rotated by 30° (red) and 15° (green) with respect to the FeRh substrate. In µ-diffraction patterns at other places several orientations of graphene domains different from the preferential ones could be identified (not shown). The diffraction ring in Fig. 6a is formed by domains much smaller than 185 nm, which is in agreement with the distribution of domain sizes observed by STM. Fig. 6e-f (right) shows a real-space schematic illustration of two preferential orientations of the graphene lattice with respect to the FeRh substrate, which correspond to the diffraction patterns in Fig. 6e. In this schematics, only the relative rotation of the graphene lattice with respect to the FeRh lattice is relevant, we cannot determine the exact relative positions of graphene atoms on the FeRh substrate. The relative rotation of the selected graphene orientations by 15° leads to a large change in orientation and spacing of the lines where carbon and rhodium atom rows overlap (indicated by blue dashed lines in Fig. 6f). Spacing of these lines corresponds to the major periodicity of the line moiré in Fig. 5b (~6 Å) and Fig. 5c (~8.4 Å).

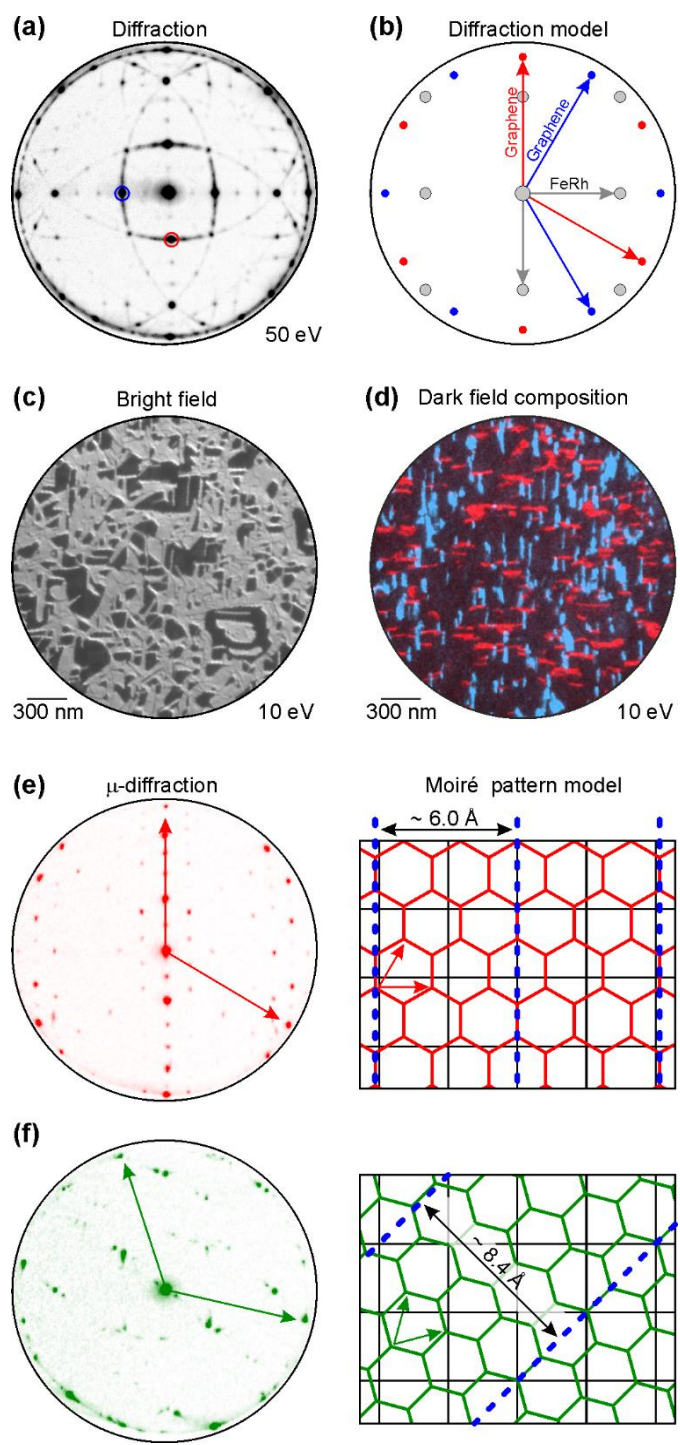

**Fig. 6.** (a) Diffraction pattern of the graphene layer on FeRh measured by LEEM at an electron energy of 50 eV. Electrons were collected from an area of 10×15 µm². The diffraction ring with 12 intense spots at the edge of the diffraction pattern corresponds to polycrystalline graphene with several preferential orientations. (b) Schematic illustration of the diffraction pattern of FeRh and two preferential graphene orientations corresponding to (a). Gray, red and blue arrows represent the reciprocal space unit vectors of the FeRh substrate and two graphene domains, respectively. (c) Bright field LEEM image showing the graphene coverage of the surface. Graphene (prepared from ethylene) corresponds to the light areas. (d) Dark field LEEM images of the spatial distribution of preferential

graphene orientations highlighted in (b) visualized by selecting the related diffraction spots in (a). (e) – (f) left, diffraction patterns measured with a µ-diffraction aperture at two different positions featuring the preferential graphene orientations rotated by 30° (red) and 15° (green) with respect to the FeRh substrate. Electrons were collected from a circular area with a diameter of 185 nm defined by the aperture. (e) – (f) right, schematics of the preferential orientations of graphene domains with respect to the FeRh atoms, which were deduced from the µ-diffractions in (e) – (f) left. In this schematics, only the relative rotation of the graphene lattice is relevant. The indicated lattice vectors correspond to the reciprocal unit vectors displayed in (e) – (f) left. The schematics of epitaxy assumes graphene on a Rh-terminated surface and lattice parameters of 2.99 Å for FeRh (Rh-Rh spacing) and 2.46 Å for graphene. The periodic lines where carbon and rhodium atom rows overlap are indicated by blue dotted lines.

The quality and homogeneity of the graphene layer is affected by several factors [49-51] controlling the processes of hydrocarbon decomposition and carbon segregation on metal substrates. Our results are in good agreement with the previously cited literature: the graphene layer on FeRh surface is either grown by segregation of carbon atoms dissolved in the FeRh film or by reduction of hydrocarbons on the FeRh surface (see also Section 2). Besides the well-controlled ethylene source the hydrocarbons originate in the airborne contamination and residual atmosphere in vacuum chambers (except for ultra-high vacuum conditions) forming adventitious carbon coverage on the surface of thin films. In both cases, the Rh terminated surface we observe after annealing of the FeRh thin film [31] acts as a catalyst for reduction of hydrocarbons creating a carbon source for graphene growth.

In general, high solubility of carbon in the substrate usually leads to formation of multilayer and polycrystalline graphene [52-54], which may be partially circumvented by the use of alloys combining metals with high carbon solubility and low carbon solubility, such as the Ni-Cu alloy [55]. This property is provided by the FeRh multilayer structure as well, limiting the carbon content by Rh layers in the ordered stack [56].

**Conclusion**

In conclusion, we prepared single-layer graphene on epitaxial FeRh thin films featuring the first-order metamagnetic phase transition. The quality of the graphene layer was characterized by high-resolution XPS, linear dichroism using NEXAFS at the C 1s absorption edge, LEIS, STM, and LEEM. Both graphene prepared by carbon segregation and graphene formed by ethylene decomposition generally displayed very similar characteristics, only in case of x-ray linear dichroism we observed more prominent peaks for the ethylene graphene. The graphene layer is tiled into domains showing distinct atomic-like contrast caused by specific orientations of graphene grains on the highly ordered epitaxial FeRh substrate. The graphene grains with a typical size of 5 – 50 nm are strongly attached to the substrate and contain freestanding graphene bubbles. LEEM confirms growth of graphene with several preferential azimuthal orientations of graphene domains.

The homogeneity of graphene was evaluated using LEIS. The graphene layer on top of FeRh covers more than 97% of the surface area and is free of significant defects. No other elements (contaminants) were detected. The graphene protected surfaces exposed to atmosphere do not show extra signatures related to air molecules. After months spent in air the clean graphene surface can be renewed with simple annealing process at 500 K which desorbs the air contamination and allows for surface sensitive experiments such as XPS, STM, LEEM and LEIS. The electronic properties of the FeRh layers are not modified in these conditions. These findings open the way to vertical stacking of 2D materials and possibly controlling their magnetic properties on tunable magnetic substrates, whose magnetic state can be controlled by temperature, magnetic field, electric field, or strain due to their multiferroic nature.

**Acknowledgments**

We thank Prof. Peter Varga for discussion on the graphene domain contrast in STM. V.U. and J.A.A. acknowledge the Grant Agency of the Czech Republic (grant no. 16-23940Y). Access to the CEITEC Nano Research Infrastructure was supported by the Ministry of Education, Youth and Sports (MEYS) of the Czech Republic under the projects CEITEC 2020 (LQ1601) and CzechNanoLab (LM2018110). P.P. and J.C. acknowledge the project TC17021 of the Inter-Excellence program of MEYS. S.P., M.P. and T.S. acknowledge the support from the H2020 Twinning program (project SINNCE, 810626) and Technology Agency of the Czech Republic (grant No. TE01020233). V.U. was supported by funding from the European Union's Horizon 2020 research and innovation program under the Marie Skłodowska-Curie that is co-financed by the South Moravian Region under grant agreement No. 665860. This project has received funding from the EU-H2020 research and innovation program under grant agreement No 654360 having benefitted from the access provided by CNRS to the SOLEIL Synchrotron, within the framework of the "NFFA-Europe Transnational Access Activity".

# Supplementary Material

# Single-layer graphene on epitaxial FeRh thin films

Vojtěch Uhlíř[a,b,*], Federico Pressacco[c], Jon Ander Arregi[a], Pavel Procházka[a,b], Stanislav Průša[a,b], Michal Potoček[a,b], Tomáš Šikola[a,b], Jan Čechal[a,b], Azzedine Bendounan[d] and Fausto Sirotti[d,e]

[a]CEITEC BUT, Brno University of Technology, Purkyňova 123, 612 00 Brno, Czech Republic

[b]Institute of Physical Engineering, Brno University of Technology, Technická 2, 61669 Brno, Czech Republic

[c]Center for Free Electron Laser Science, University of Hamburg, Luruper Chaussee 149, 22761 Hamburg, Germany

[d]Synchrotron SOLEIL, Saint-Aubin, BP 48, F-91192 Gif-sur-Yvette Cedex, France

[e]Physique de la Matière Condensée, CNRS and École Polytechnique, Université Paris Saclay, F-91128 Palaiseau, France

*Email: vojtech.uhlir@ceitec.vutbr.cz

**Magnetization vs. temperature characterization of as-grown FeRh films**

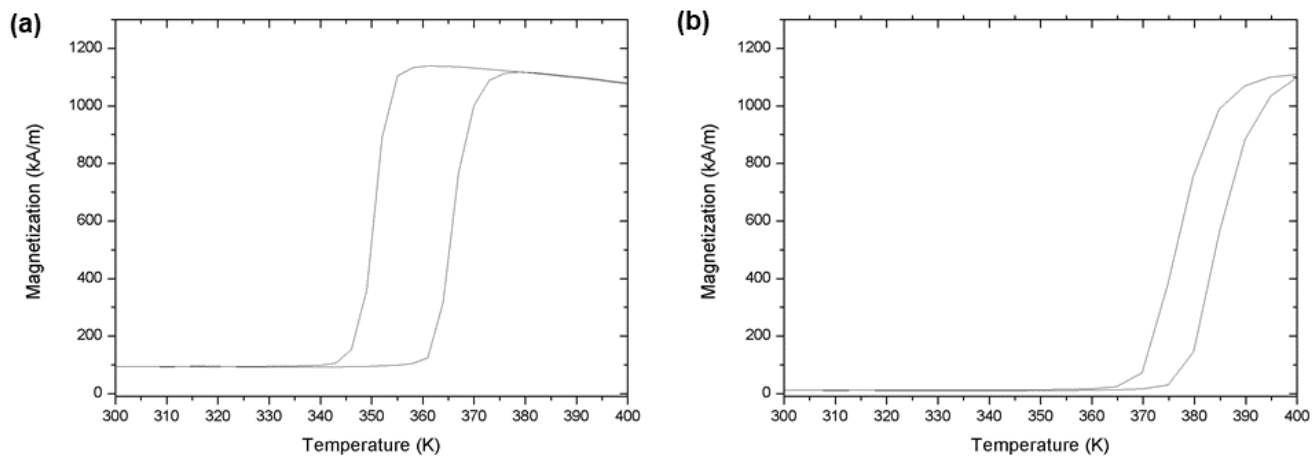

**Fig. S1.** Magnetization as a function of temperature showing the metamagnetic phase transition in a 55-nm-thick FeRh film (a) and 150-nm-thick film (b). Characterization was done in magnetic field of 1 T, which offsets the phase transition by 8 K to lower temperatures.

**Depth profiling of carbon segregation in FeRh thin films by SIMS**

The carbon content in FeRh thin films at different stages of heat treatment was determined by depth profiling using secondary ion mass spectroscopy (SIMS) with a time-of-flight analyzer (TOF-SIMS 5 instrument from Iontof). We have used the co-sputtering interlaced mode with $Bi_1^+$ and $Cs^+$ ion beams. The depth sputtering was done with $Cs^+$ ions (beam energy 500 eV) in a 300 × 300 µm² area and the analysis in a 200 × 200 µm² area was done by $Bi_1^+$ ions (beam energy 30 keV) in the spectroscopy mode. The mass resolution was approximately 7000.

The carbon depth profiles of 40-nm-thick FeRh films for different conditions of heat treatment and doping are shown in Fig. S2. All signals were calibrated to the Rh signal obtained from depth profiling of the FeRh film, which shows a homogeneous Rh composition across the film thickness. A FeRh thin film prepared according to a standard protocol for obtaining well-ordered epitaxial films, i.e. annealing in situ after deposition at 1070 K (see Experimental Section), is represented by the black profile. The carbon concentration is homogeneous across the film thickness.

We have compared three 40-nm-thick FeRh films deposited at room temperature, which were doped (co-deposited) with 0%, 4%, and 9% of carbon (respectively red, blue, and green signals in Fig. S2a). Regarding the scaling of the signal with carbon concentration, we compare the reference films of 4% doped C and 9% doped C. The ratio of the carbon signal in these films is 2.45, which is in agreement with the ratio of the C doping in these films. We can then estimate the amount of carbon in the annealed film with 0% doped C to approximately 0.7 at% C.

In case of room temperature deposited FeRh thin films, we observe a carbon peak at the interface with the MgO substrate, which comes from airborne contamination. These carbon-based molecules are typically removed by preheating the substrate at 720 K prior to the FeRh deposition. In any case, intensity of this interfacial peak rapidly decreases across the depth profile upon subsequent annealing cycles (Fig. S2b-c).

The depth profiles of the FeRh film originally deposited and annealed at elevated temperatures are indicated by gray and black lines in Fig. S2b. Two subsequent heating cycles are applied, each one corresponding to 10 min at 870 K. We observe a gradual decrease of the carbon amount after each of the annealing cycles, down to approximately 0.1 at% C. A carbon-depletion zone becomes apparent close the FeRh film surface. However, there is an increase of the carbon content at the surface of the FeRh film, which unfortunately cannot be quantified due to insufficient number of points in the profile there.

The FeRh film doped with 4 at% C (light blue and dark blue lines in Fig. S2c) shows a clear trend of carbon segregation towards the FeRh surface upon repeated heating cycles. The interfacial carbon is leveled out upon the first annealing cycle.

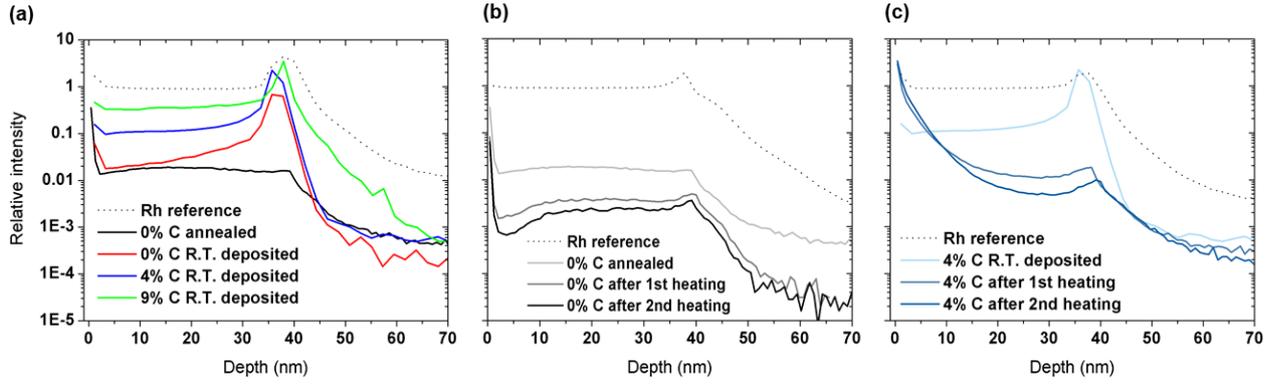

**Fig. S2.** (a) Depth profiles of C concentration in FeRh thin films subjected to different initial heat treatment and C doping. (b)-(c) Segregation of C at the FeRh surface and reduction of C in the film upon repeated heating cycles.

Finally, we estimate the minimum FeRh film thickness containing sufficient carbon to form a single graphene layer. The number of carbon atoms needed to form a single layer of graphene amounts to $N_C = n_G S$, where $n_G$ is the area concentration of C atoms in graphene (2 atoms per unit cell, $S_{cell} = 5.24$ Å$^2$), and $S$ is the sample area. The number of carbon atoms in a FeRh film equals $N_C = n_C S t$, where $n_C$ is the volume concentration of C atoms in a FeRh unit cell (1 Fe and 1 Rh atom per unit cell, $V_{cell} = 26.7$ Å$^3$):

$$n_C = \frac{at\% \, C}{100} \cdot \frac{N_{FeRh}}{V_{cell}},$$

and $t$ is the film thickness. Using the known parameters for graphene and FeRh we obtain a formula for the minimum thickness of the FeRh film with a defined carbon concentration to form single-layer graphene:

$$t = \frac{n_G}{n_C} = \frac{26.7 \cdot 10^{-30}}{5.24 \cdot 10^{-20}} \cdot \frac{100}{at\% \, C} \, m \cong \frac{51 \, nm}{at\% \, C}.$$

Hence, for the carbon concentration of 0.7 at% we find in the originally annealed FeRh films, 73 nm of FeRh would contain enough carbon to form one single layer of graphene over the whole sample area.

### Identification of preferential graphene orientations in the LEED patterns

The four dominant spots in Fig. S3a are assigned to the square lattice of the FeRh substrate. They are indicated by red dots in a calculated diffraction pattern in Fig. S3b. The ring close to the edge of the diffraction pattern in Fig. S3a with a radius matching the reciprocal lattice parameter of graphene is a clear sign of a polycrystalline character of the graphene layer with many different grain orientations.

We recognize several preferential graphene orientations giving sharp spots around the graphene ring. The most prominent ones are marked by purple dots in Fig. S3b. Here, the rotation of graphene reciprocal unit vector $\vec{a}_G^*$ with respect to the FeRh reciprocal unit vector $\vec{a}_{FeRh}^*$ is 0°. Taking into account the graphene hexagonal lattice and four-fold rotation symmetry of the substrate we obtain 12 symmetry equivalent diffraction spots associated with two possible symmetry equivalent graphene orientations. In real space the corresponding rotation of the graphene unit vector $\vec{a}_G$ with respect to the FeRh unit vector $\vec{a}_{FeRh}$ is ±30°. The spots associated with the second most prominent orientation are rotated by ±45° (blue) associated with two symmetry equivalent graphene orientations with the real space graphene rotation by ±15° (blue) with respect to $\vec{a}_{FeRh}$. We note that some of these spots coincide with (1,1) spots of the FeRh substrate. Finally, slightly weaker spots at -49°, 41°, -41°, and 49° (green) are associated with graphene rotations by ±11° and ±19°, all of which are symmetry equivalent with respect to the FeRh substrate.

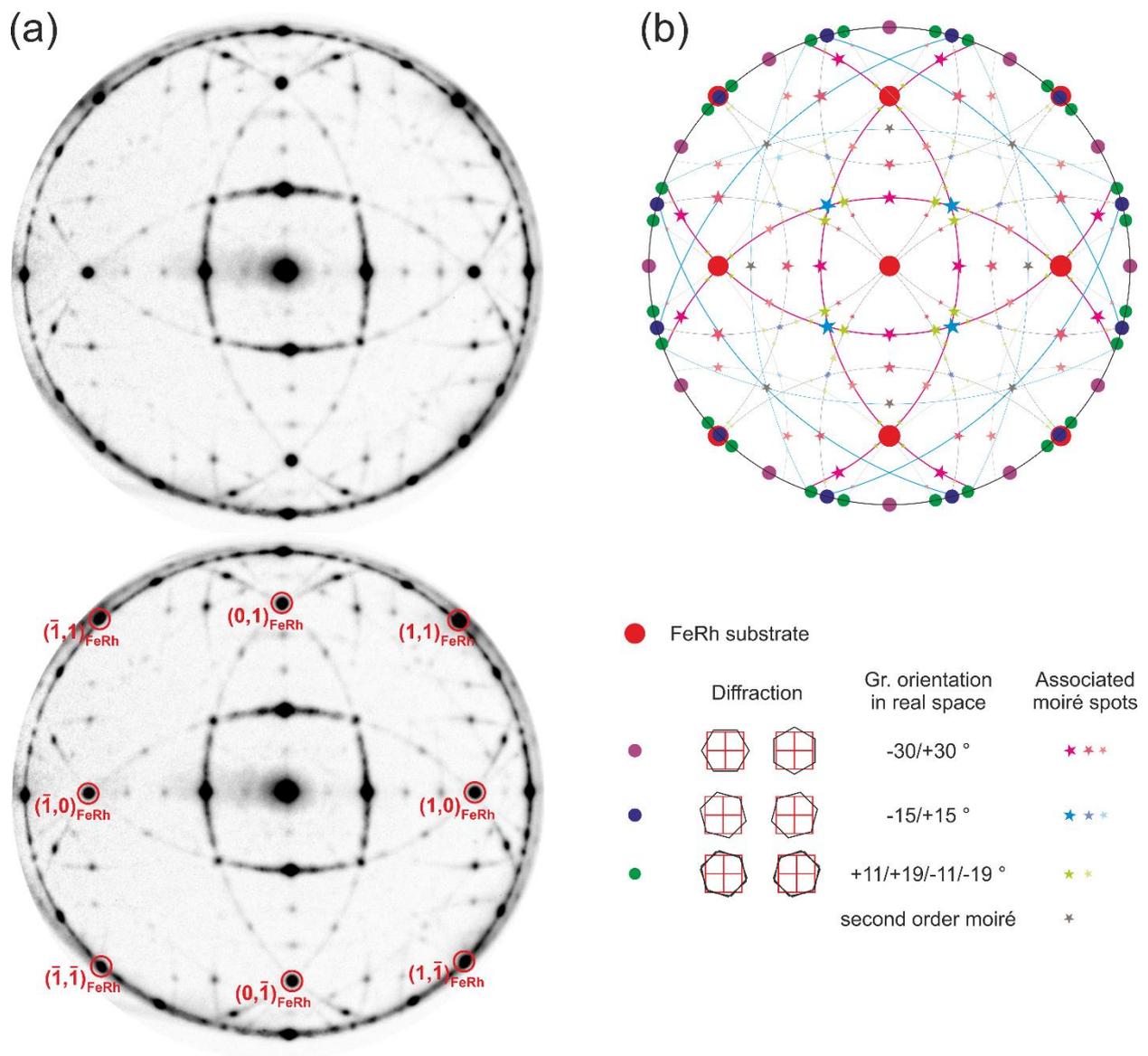

**Fig. S3.** (a) LEED pattern with indexing of the FeRh spots indicated in the bottom panel. (b) Associated model of the diffraction and moiré spots and lines of the preferential graphene orientations.

**Moiré spots and lines**

Fig. S3b presents calculated moiré spots and lines in the diffraction pattern. The moiré spots are calculated employing geometric construction [1]. First, the reciprocal moiré unit vector is obtained as $\vec{K}_{\text{moiré}} = \vec{a}_G^* - \vec{a}_{\text{FeRh}}^*$. Subsequently, moiré spots are obtained as a linear combination of reciprocal lattice of the substrate with $\vec{K}_{\text{moiré}}$. Taking $a_{\text{FeRh}}$ = 2.99 Å, $a_G$ = 2.46 Å and calculating the first order moiré spots for continuously rotating $\vec{a}_G$ with respect to $\vec{a}_{\text{FeRh}}$ we reproduce the line features in the diffraction pattern. Considering the double periodicity ($2\vec{K}_{\text{moiré}}$) we also reproduce the light blue lines of a double radius. For the preferential graphene orientations sharp moiré spots laying on the moiré lines are visible in Fig. S3a: these are represented by stars in the calculated pattern in Fig. S3b. The diffraction pattern is thus a clear signature of presence of graphene domains in random rotations with respect to the substrate with several preferential rotations giving rise to discrete diffraction patterns.